\documentclass[10pt]{iopart}

\usepackage{graphicx}
\usepackage{hyperref}
\usepackage{lineno, color}
\usepackage{xcolor}
\begin{document}
\hypersetup{
pdftitle={Mitigation of plasma-wall interactions with low-Z powders in DIII-D high confinement plasmas},
pdfsubject={plasma boundary research},
pdfauthor={Florian Effenberg},
pdfkeywords={divertor power exhaust, dust injection, wall conditioning, detachment, powder injection, impurity seeding, core-edge integration}
}

\title[Mitigation of plasma-wall interactions with low-Z powders in DIII-D high confinement plasmas]{Mitigation of plasma-wall interactions with low-Z powders in DIII-D high confinement plasmas}

\author{F. Effenberg$^{1}$, A. Bortolon$^{1}$, L. Casali$^{2}$, R. Nazikian$^{3}$, I. Bykov$^{3}$, F. Scotti$^{4}$, H.Q. Wang$^{3}$,  M.E. Fenstermacher$^{4}$, R. Lunsford$^{1}$, A. Nagy$^{1}$, B.A. Grierson$^{3}$, F. M. Laggner$^{1}$, R. Maingi$^{1}$, and the DIII-D Team}

\address{1 - Princeton Plasma Physics Laboratory, Princeton, NJ 08543, USA}
\address{2 - University of Tennessee, Knoxville, TN 37996, USA}
\address{3 - General Atomics, San Diego, CA 92186, USA}
\address{4 - Lawrence Livermore National Laboratory, Livermore, CA 94550, USA}

\ead{feffenbe@pppl.gov}
\vspace{10pt}
\begin{indented}
\item[]Submitted to Nuclear Fusion, March 2022
\end{indented}

\begin{abstract}
Experiments with low-Z powder injection in DIII-D high confinement discharges demonstrated increased divertor dissipation and detachment while maintaining good core energy confinement. Lithium (Li), boron (B), and boron nitride (BN) powders were injected in H‑mode plasmas ($I_p=$1 MA, $B_t=$2 T, $P_{NB}=$6 MW, $\langle n_e\rangle=3.6-5.0\cdot10^{19}$ m$^{-3}$) into the upper small-angle slot (SAS) divertor for 2-s intervals at constant rates of 3-204 mg/s. 
 
The multi-species BN powders at a rate of 54 mg/s showed the most substantial increase in divertor neutral compression by more than an order of magnitude and lasting detachment with minor degradation of the stored magnetic energy $W_{mhd}$ by 5\%. Rates of 204 mg/s of boron nitride powder further reduce ELM-fluxes on the divertor but also cause a drop in confinement performance by 24\% due to the onset of an $n=2$ tearing mode.

The application of powders also showed a substantial improvement of wall conditions manifesting in reduced wall fueling source and intrinsic carbon and oxygen content in response to the cumulative injection of non-recycling materials.

The results suggest that low-Z powder injection, including mixed element compounds, is a promising new core-edge compatible technique that simultaneously enables divertor detachment and improves wall conditions during high confinement operation.
\end{abstract}
%
\vspace{2pc}
\ioptwocol
\section{Introduction}
Present plasma-facing component (PFCs) materials cannot withstand continuous heat fluxes above 10 MW/m$^2$ predicted for the next step large-scale fusion reactors such as ITER \cite{pitts_full_2013} or a projected future US fusion pilot plant \cite{national_academy_of_engineering_bringing_2021}. Dissipation and spreading of heat fluxes by isotropic low-Z impurity line emission are promising techniques to prevent melting and damage and extend the lifetime of the exposed wall materials. Usually, low-Z impurities are injected in gaseous form to enhance divertor radiation or create a radiative mantle surrounding the hot core plasma \cite{maddison_moderation_2011}. Impurity gas seeding has been developed for this purpose and extensively investigated, e.g., at the tokamaks TEXTOR, JET, Alcator C-Mod, ASDEX Upgrade, DIII-D, EAST, the heliotron LHD, and more recently at the stellarator W7-X \cite{samm_radiative_1993, maddison_moderation_2011, reinke_2011, kallenbach_partial_2015, casali_improved_2020, chen_2017, morisaki_radiated_2015, effenberg_first_2019}. 

The radiative mantle and divertor dissipation approach prevent damage and wear of main plasma-facing components to enable safe long-pulse operation. However, the ideal solution must be core-edge compatible, i.e., power losses in the boundary and divertor plasma must not have a detrimental impact on the high-performance core plasma.

The radiative power losses by low-Z impurities reduce the temperature and sputtering in the divertor. Plasma recombination sets on at temperatures below 1 eV and creates a protective neutral gas cushion in front of the targets, protecting against incoming heat and particle fluxes. In such a regime, the plasma boundary is removed from the divertor targets, and erosion and sputtering processes are substantially suppressed. This state is called plasma detachment \cite{matthews_plasma_1995, krasheninnikov_2016}.

Nitrogen and neon gases are widely used for effective power dissipation in the plasma boundary. However, nitrogen is of concern in a reactor environment due to the production of tritiated ammonia (NT3), which contaminate the uranium beds of the ITER tritium plant \cite{neuwirth_2012,laguardia_2017}. The choice of impurities is eventually determined by their radiation characteristics and may be limited by chemistry \cite{kallenbach_impurity_2013, walker_neutron_2017}. 

Numerical studies have suggested that alternative impurities such as boron (B) deserve consideration for radiative dissipation as well \cite{pigarov_radiative_2017, effenberg_2020}. However, boron delivery in gaseous form requires toxic or explosive gases like diborane (B$_2$D$_6$) or strongly diluted compounds such as trimethyl borane (C$_3$D$_9$B).

Lithium is strongly considered a liquid metal wall candidate, potentially allowing to spread and conduct heat efficiently and safely away \cite{poradzinski_2019, andruczyk_2020, marenkov_2021}. Studies also predict it to be an effective dissipator in the lithium vapor box divertor \cite{goldston_lithium_2016, schwartz_2019}. 

More recently, the injection of low-Z solid materials in powder, dust, or small granule form was introduced as a novel technique for various real-time applications. 
First applications focused on real-time wall conditioning, disruption mitigation, mitigation of edge localized modes (ELMs), and confinement improvement with boron, boron nitride (BN), and lithium powders at TFTR, T-10, DIII-D, EAST, ASDEX Upgrade, KSTAR, LHD, and W7-X \cite{snipes_1992, sergeev_2012, bortolon_real-time_2019, bortolon_observations_2020, gilson_wall_2021, sun_real_2019, maingi_elm_2018, sun_suppression_2021, nespoli_2020, lunsford_characterization_2021}. Real-time injection of powders has been particularly beneficial for long-pulse operation and at experiments using superconducting coils, where it can supplement or even substitute conventional glow discharge boronization. Low Z material injection is also being considered a method for replenishing wall claddings and surface conditioning in fusion pilot plants \cite{stangeby_2022}.

In the following, first-time experiments at DIII-D are presented that show strong divertor dissipation and mitigation of plasma-wall interactions during H-mode by injecting lithium, boron, and boron nitride powders into the upper closed small angle slot divertor. 

The remainder of this paper is organized as follows: Section \ref{sec:setup} provides an overview of the experiment and method of impurity injection; sections \ref{sec:powders_I} and \ref{sec:powders_II} present the effect of lithium, boron, and boron nitride injection on confined and divertor plasma; section \ref{sec:RTWC} illustrates the beneficial effect of real-time low-recycling powder injection on wall-conditions; section \ref{sec:discussion} discusses the results in the context of recent studies; the final section \ref{sec:summary} provides a summary.

\begin{figure}
\begin{center}
\includegraphics[width=0.48\textwidth]{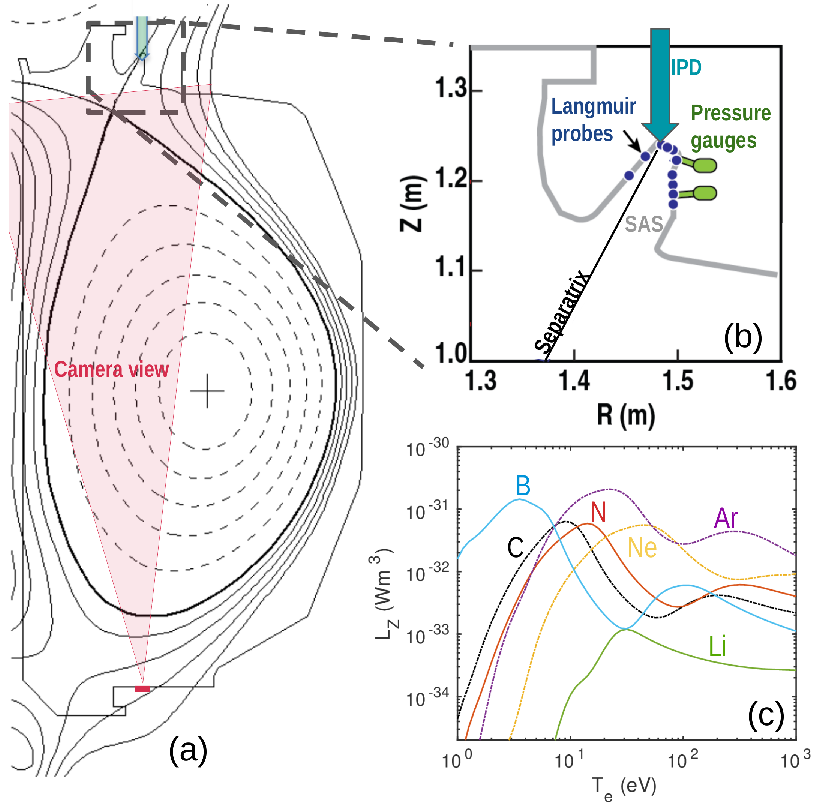}
\caption{\label{fig:Figure1} (a) Magnetic equilibrium and of upper single null H-mode scenarios. The arrow indicates the powder injection location in the small-angle slot divertor (SAS). (b) Langmuir probes and pressure gauges, impurity powder dropping location in the SAS divertor. (c) Power loss function $L_Z=L_Z(T_e,n_e\tau)$ for lithium, boron, carbon, nitrogen, neon, and argon in non-coronal equilibrium for $n_e\tau=10^{20}$ m$^{-3}$ms.}
\end{center}
\end{figure}

\section{Setup of low-Z powder injection in DIII-D high confinement plasmas}\label{sec:setup}
The experiments were conducted in upper-single null ELMy H‑mode in standard and reversed toroidal field ($B_t$) direction ($I_p=1.0$ MA, $B_t=2$ T, $P_{NB}=6$ MW, $\beta_n=2.0$, $f_{ELM}=80$ Hz, $n_e=3.6-5.0\cdot$10$^{19}$ m$^{-3}$). The outer strike point (OSP) is positioned in the small-angle slot divertor. In this configuration, the ion grad-B ($B\times\nabla B$) drift direction points away from (into) the SAS divertor in case of standard (reversed) toroidal B$_t$ direction. Figure \ref{fig:Figure1}(a) shows the magnetic equilibrium, the location of the OSP at the SAS target, the powder injection location (green arrow),  and the vertical camera view (DiMES TV \cite{abrams_2017}). The SAS geometry, the array of Langmuir probes (LP), the position of pressure gauges (PG), and the injection location of the impurity powder dropper (IPD) are shown in figure \ref{fig:Figure1}(b) in more detail. The divertor Langmuir probes and pressure gauge are used to characterize the divertor plasma. Vertical and tangential camera diagnostics measure visible and filtered line integrated brightness of different charge states of injected impurities \cite{fenstermacher_tangentially_1997, abrams_2017}. The Multi-chord Divertor Spectrometer (MDS) provides additionally high-resolution spectroscopy for the divertor SOL \cite{brooks_1992}.

The closed divertor geometry generally facilitates the dissipation of heat and particle fluxes due to inherently better neutral compression \cite{morisaki_2013, guo_first_2019, shafer_dependence_2019, casali_2019, casali_2021, fevrier_2021}.
 
The species-dependent cooling behavior of different elements used in dissipation experiments is usually expressed by their radiative losses $P_{loss,rad}=n_e n_Z L_Z$, which are proportional to the electron density $n_e$, the impurity density $n_Z$ and the power loss function $L_Z$. The power loss function describes the cooling efficiency for an impurity as a function of only the electron temperature in coronal equilibrium in the core plasma \cite{kallenbach_2019}. However, the plasma edge is in non-coronal equilibrium due to steep gradients, and associated transport \cite{carolan_1983}. Here, the power loss function also depends on the non-equilibrium parameter $n_e\tau$ with $\tau$ being the particle residence time: $L_Z=L_Z(T_e,n_e\tau)$.  
The non-equilibrium power loss function is shown in figure \ref{fig:Figure1}(c), assuming $n_e=10^{20}$ m$^{-3}$ and a particle residence time of 1 ms in the plasma edge for lithium, boron, carbon, nitrogen, neon, and argon. These elements are introduced in powder form in the present study (B, Li, BN) or are intrinsically released (C) through erosion or extrinsically injected as impurity gas (N, Ne, Ar) during conventional radiative power exhaust experiments. The total line emission has been calculated with atomic data from ADAS \cite{summers_atomic_2002}. Boron shows for $L_Z$ a maximum at a few eV while lithium reaches its maximum at a few tens of eV corresponding to far and near scrape-off layer (SOL) temperatures, respectively. Nitrogen shows peak radiation around 10 eV. The radiation potentials suggest that B, N, C causes power losses mainly in the SOL, while Li and Ne may radiate in the near SOL and around the separatrix. Ar shows substantial losses also at higher temperatures in the confinement.
\begin{figure}
\begin{center}
\includegraphics[width=0.48\textwidth]{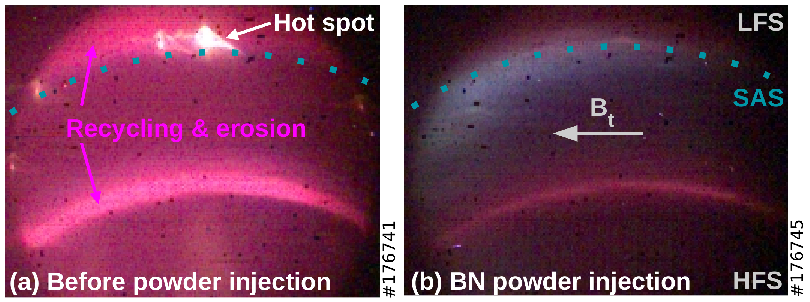}
\caption{\label{fig:Figure2} Vertical bottom-up view into closed divertor by visible camera (figure \ref{fig:Figure1}(a)) imaging (a) prior and (b) during boron nitride powder injection. The toroidal field direction ($B_t$), low field side (LFS), and high field side (HFS) are marked. The dashed line marks the position of the small angle slot (SAS) divertor. The bright emission belts and spots represent boundary emission from plasma-wall interactions and hot spot spots.}
\end{center}
\end{figure}
The impurity powder dropper (IPD) \cite{nagy_multi-species_2018} is mounted at the top of the DIII-D tokamak device at a toroidal angle of 195$^o$. The radial drop location is $R=1.484$ m. It has four reservoirs filled with different types of impurity powders. Impurity powders were dropped directly into the OSP region through a tube of $2$ $m$ length into the SAS. The powder particles reach velocities of $5-6$ $m/s$ upon divertor entrance due to acceleration by gravitational force. The time delay between actuation and impact on the divertor plasma with the current DIII-D setup is typically 900 ms, whereas impurity gas feedback allows response within 250-500 ms. The macroscopic powder particles must undergo ablation, which depends on the powder particle size. Lithium and boron nitride powders were of relatively fine size (up to $40$ and $65$ $\mu m$), while the boron powder used consists of relatively large sizes (up to $150$ $\mu m$).

\section{Cooling of the plasma boundary with lithium, boron and boron nitride powders} \label{sec:powders_I}
In a sequence of experiments, boron, boron nitride, and lithium were injected in powder form in H-mode (\#176740-52). The low-Z powders were released into the divertor plasma at ~2.9 s. Mass flow rates of 3-204 mg/s were determined with a flow meter \cite{nagy_multi-species_2018} corresponding to atomic rates of $0.2 -5.4\cdot 10^{21} \frac{atoms}{s}$. 

A bottom-up camera view (figure \ref{fig:Figure1}(a)) was used to capture the visible emission from plasma-wall interactions on the high field side (HFS) and low field side (LFS) plasma-facing components. The emission is shown in figure \ref{fig:Figure2}(a) before and (b) during BN powder injection of experiments conducted in the standard configuration with the ion grad-B drift pointing out of the upper SAS divertor. Before powder injection, bright emission was present at and near the outer and inner strike lines and the recycling regions (figure \ref{fig:Figure2}(a)). Also, a hot spot was detected on the low field side at a leading tile edge. The strong boundary emission at the low and high field sides results from neutral $D_{\alpha}$ and intrinsic impurities sputtered from the plasma-facing components. During injection of BN powder ($\geq200$ mg/s), boundary emission and hot spot were suppressed (figure \ref{fig:Figure2}(b)). 

\begin{figure}
\begin{center}
\includegraphics[width=0.48\textwidth]{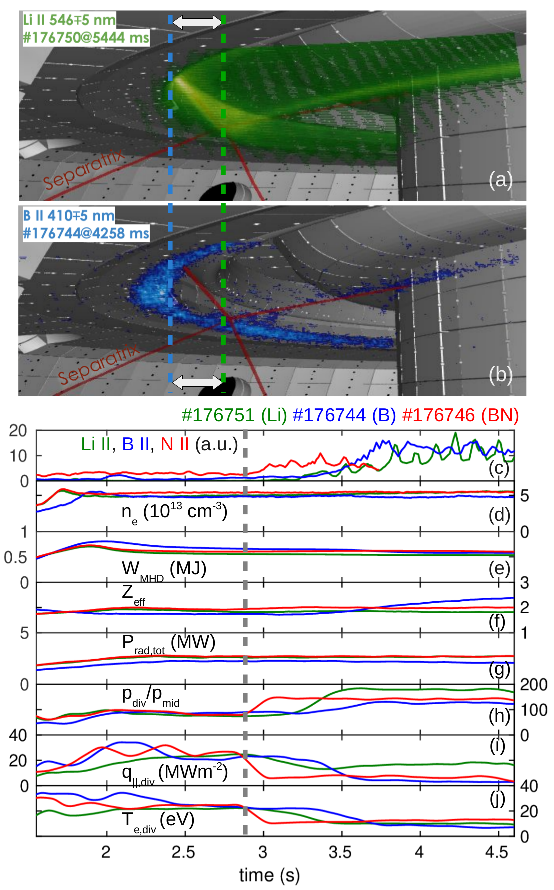}
\caption{\label{fig:Figure3} Line emissivities for (a) Li II (546 nm) and (b) B II (410 nm) were observed with tangential cameras. Vertical dashed lines indicate the innermost radial extend of the B II and Li II radiation zones. Time traces of (c) divertor spectral line emissivities of Li II (546 nm), B II (608 nm) and N II (399 nm), (d) line averaged electron density, (e) stored magnetic energy $W_{MHD}$, (f) effective charge ($Z_{eff}$), (g) total radiated power $P_{rad,tot}$, (h) neutral compression (ratio of divertor-to-midplane pressure), (i) peak divertor parallel heat flux $q_{\parallel, div}$, and (j) peak divertor electron temperature $T_{e,div}$. Lithium (3.3 mg/s), boron (35 mg/s) and boron nitride (13 mg/s) powder injections start at $\approx 2.8-2.9$ seconds.}
\end{center}
\end{figure}

During experiments with lithium and boron powder injections (3-35 mg/s), components of the spatial emission distributions are captured with the tangential camera. The camera filters extract Li II at 546 nm and B II at 410 nm, which are the only proxies for the spatial distribution of B and Li radiative power losses at this point. Since D$_{\delta}$ and C III emit within the band-pass of the filter used to image boron emission, background emission before boron injection was subtracted from the camera images. However, some residual contamination might remain due to a change of emissivity resulting from local cooling from the injected powder. The spatial brightness distributions are shown in figure \ref{fig:Figure3} for single ionized lithium (a) and boron (b) at times during the injection, i.e., 5444 ms and 4258 ms, respectively. Vertical blue and green dashed lines in figure \ref{fig:Figure3}(a) and (b) indicate the innermost radial extend of the B II and Li II radiation zones. The Li II radiation is concentrated along the outer divertor leg, extending from the SAS volume to the X-point. The B II emission shows stronger brightness in a layer close to the PFCs and extends in the radial direction outwards near the targets. There is almost no overlap in the spatial distributions between the Li II and B II radiation in the radial direction. The distinct features of the emission distribution of Li and B are consistent with the temperature dependency of the cooling potentials (figure \ref{fig:Figure1}(c)). Boron cools the SOL while lithium causes temperature reduction in the vicinity of the separatrix. Unfortunately, during these first-time experiments, matching the powder flow rates between different species was not possible due to a lack of fine control. However, the results show substantial cooling of the divertor plasma with $3.3$ $mg/s$ lithium and $35$ $mg/s$ boron powder injection. 
The time traces in figure \ref{fig:Figure3}(c-j) show a comparison of three representative H-mode plasmas with lithium, boron and boron nitride powder injections. The Li II (547 nm), B II (608 nm), and N II (399 nm) line emissivities measured with the Multi-chord Divertor Spectrometer increase when the powder reaches the plasma around 2.8-2.9 seconds (however, data for N II are missing after 3.7 seconds). The line averaged density increases only slightly by 5\% during lithium injection, drops by 8\% during boron injection and remains constant for BN injection. $Z_{eff}$ remains constant in the case of Li, and nearly constant in the case of BN ($\Delta Z_{eff}\approx+3\%$) but increases from 1.8 to 2.4 during B injection. The stored magnetic energy reduces by 5\% in the case of Li and remains unchanged during BN injection. In the case of boron injection, a rate of $35$ $mg/s$ eventually triggers a $n=2$ tearing mode which reduces stored magnetic energy by 14\% after 3.5 s. The total radiated power remains mostly stable at 2.6 MW for Li and BN and 2.2 MW for B injection. The coverage of the small angle slot divertor is insufficient for quantifying the divertor radiation. The signals from individual bolometer chords closest to the SAS divertor suggest an increase in local radiation by 15-30\% during Li and B injection, which does not significantly alter the measured total radiated power $P_{rad,tot}$ (figure \ref{fig:Figure3}(g)).
Dilution effects are estimated by measuring the rate of neutrons produced by deuterium-deuterium collisions during neutral beam injection. The measured neutron rates are related to the expected neutron emission based on a zero-dimensional calculation that includes the carbon density as the only impurity source \cite{heidbrink_1997} which allows estimating the dilution effects of injected impurities. The neutron rates before powder injection are $0.8-1.1\times 10^{15}$ n/s. Deuterium plasma dilution with Li and B impurities causes a drop of neutron rates by up to 10\% and 25\%.
The divertor neutral compression shown in figure \ref{fig:Figure3}(h) is defined as the ratio of divertor neutral pressure to main chamber pressure (measured at the outboard midplane). The divertor neutral gas pressure increases from 0.3 mtorr to 0.9 mtorr in the case of Li but only to 0.5 mtorr in the case of B injection. The divertor neutral compression increases by a factor of 3, 1.5, and 1.8 following the lithium, boron, and boron nitride powder injections. The peak electron temperatures and peak parallel heat fluxes measured with divertor Langmuir probes are shown in figure \ref{fig:Figure3}(h,i) reduce substantially after powder injection. In the case of low-rate Li powder injection, the peak electron temperature and heat flux are reduced by 12 eV and 10 $MW/m^2$, respectively. A medium rate of B results in a reduction by 20 eV and heat flux detachment. BN caused a rapid drop in $q_{\parallel,div}$ by 75\% and drop of $T_{e,div}$ from 25 eV by 15 eV within 200 ms within 200 ms. At these relatively low rates, Li enhances the particle flux and density at the peak location contributing to a partial recovery of the parallel heat flux. On the other hand, the boron and boron nitride powders also reduce the divertor particle fluxes and densities.

\section{Sustained divertor detachment and strong neutral compression with boron nitride powder}\label{sec:powders_II}
\begin{figure}
\begin{center}
\includegraphics[width=0.48\textwidth]{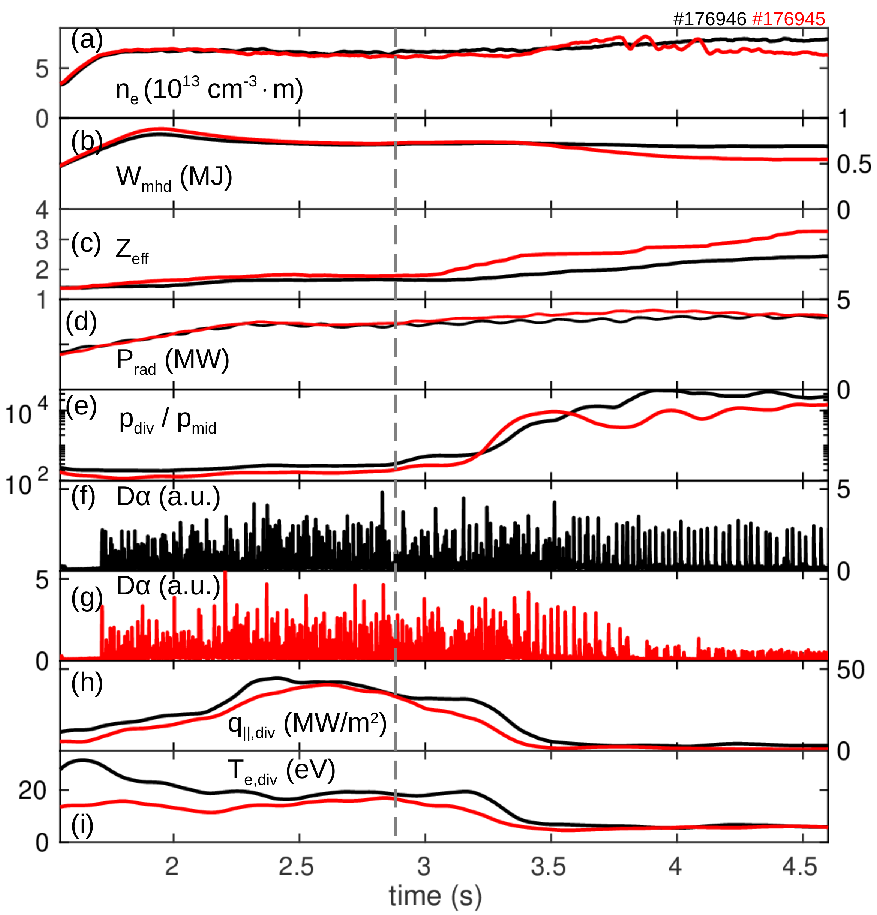}
\caption{\label{fig:Figure4} Time traces for boron nitride (BN) at high injection rates of 54 mg/s (black) and 204 mg/s (red). (a) line averaged electron density (b) stored magnetic energy $W_{mhd}$, (c) effective charge $Z_{eff}$, (d) total radiated power $P_{rad, tot}$, (e) divertor neutral compression, (f, g) divertor neutral emission $D_{\alpha}$, (h) peak divertor parallel heat flux $q_{\parallel, div}$, and (i) peak divertor electron temperature $T_{e,div}$. The powder injection starts at $\approx 2.9$ seconds.}
\end{center}
\end{figure}

Injection of boron nitride at higher rates of 54 $mg/s$ (\#176946) and above (\#176945) in configurations with ion $B\times \nabla B$ drift directed into the upper SAS divertor resulted in detachment and strong neutral compression. Figure \ref{fig:Figure4}(a-i) shows the time traces of line averaged density, stored magnetic energy, effective charge state $Z_{eff}$, total radiated power, divertor neutral compression, neutral deuterium emission in the divertor $D_{\alpha}$ and the divertor peak parallel heat fluxes and electron temperatures in the divertor. Divertor heat fluxes and temperature rapidly decrease and remain suppressed during BN injection ($T_{e,div}\leq 5$ eV). The increase in divertor neutral gas compression by more than one order of magnitude means a substantial contribution to dissipation by ion-neutral friction forces in the small-angle slot divertor. The $D_{\alpha}$ emission shows that ELM-fluxes in the divertor reduce in frequency for 54 mg/s and further reduce in amplitude at a very high rate of 204 mg/s due to dissipation.
The effective charge $Z_{eff}$ increases from base level by 0.8 to 2.3 while the stored magnetic energy $W_{mhd}$ drops from 0.7 MJ by 5\% during 54 mg/s BN injection. At the high rate of 204 mg/s, energy confinement drops by 24\% due to an $n=2$ tearing mode triggered after 3.8 s and $Z_{eff}$ increases to 4.5. The total radiated power increases by 15\%, and 25\%, respectively (figure \ref{fig:Figure4}(d)). Individual bolometer chords near the SAS divertor show a local increase in radiated power by at least 40\% and 60\%, respectively. Using the approach mentioned above, the rate of neutrons reduces by up to 28\%-38\% at higher BN powder injection rates (54-204 mg/s) due to dilution effects on the core plasma.
The rate of neutrons produced during deuterium-deuterium collisions during neutral beam injection has dropped to 62\% of the initial level ($8.8\cdot10^{14}$ n/s) at 4.5 s due to dilution effects on the core plasma during moderate and high BN injection (54-204 mg/s). However, the neutron rates (up to $1\cdot10^{15}$ n/s) were maintained without or with only more minor losses during other scenarios with low-Z powder injection. Therefore, it is suggested that the core-edge compatibility can be optimized by modifying the injection rates and powder mix for a specific plasma scenario. 

\section{Improving wall conditions} \label{sec:RTWC}
\begin{figure}
\begin{center}
\includegraphics[width=0.48\textwidth]{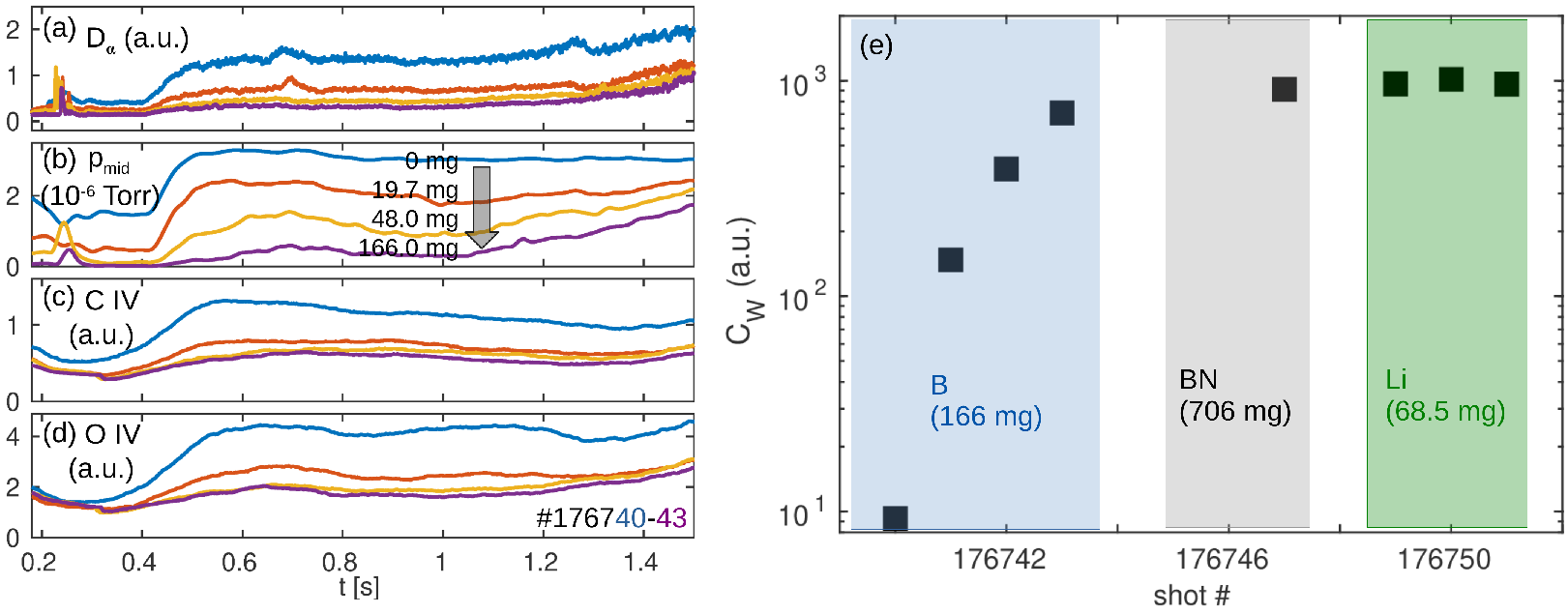}
\caption{\label{fig:Figure5} Change in wall conditioning before and after boron powder injection represented by time traces of (a) $D_{\alpha}$ line emission in the divertor, (b) main chamber neutral pressure $p_{mid}$ measured at the midplane, (c) C IV (384 nm) emission, and (d) O IV (555 nm) emission during initial L-mode phase. The cumulative mass (mg) of boron powder deposited prior to each experiment is indicated. (e) Shot-by-shot change of the ratio of deuterium gas input $\Gamma_{D2}$ to electron density $n_e$ (wall conditioning metric $C_W$), which have been calculated as an average value over a $1.2$ s time window (from $t=0.3$ s to $1.5$ s) during the initial L-mode phase. The cumulative amounts of powder are indicated in mg for each powder type.}
\end{center}
\end{figure}

Successive powder injections into the upper divertor during these experiments (\#176740-51) resulted in a significant improvement of wall conditions. The strongest effects of wall conditioning were measured during the first sequence of boron powder injection experiments for which in figure \ref{fig:Figure5}, the time traces of (a) $D_{\alpha}$, (b) main chamber neutral pressure $p_{mid}$, (c) C IV and (d) O IV are shown for the initial L-mode phase ($<1.6$ s). 

$D_{\alpha}$ and main chamber neutral pressure reduce after each discharge with B powder injection. The reduction in $D_{\alpha}$ emission and neutral pressure suggests a decrease of the intrinsic neutral source, i.e., wall recycling and out-gassing. Higher levels of neutral gas are injected to achieve the same target density following B injections. At the same time, $C IV$ and $O IV$ brightness reduce, which suggests a reduction in the intrinsic impurity source. A reduction in the impurity line emission may in parts be due to an increase in the boundary plasma temperature following a reduction in density. Unfortunately, no upstream SOL electron temperature measurements are available in this case.
The wall conditioning metric $C_W$ shown in figure \ref{fig:Figure5} (e) is defined as the ratio between deuterium gas input and line averaged electron density evaluated during the L-mode phase from $t=0.3$ s to $1.5$ s: $C_W\propto \frac{\Gamma_{D2}}{n_e}$. Improving wall conditions based on this metric means reducing the intrinsic neutral source, which allows for better plasma density control through external gas input. A better density control through improved wall conditions was achieved, expressed in the ratio between gas input and density sharply increasing during the first boron injection experiments. The wall conditions dramatically improve after injection of the first 166 mg of B: $\frac{C_W(166~ mg ~B)}{C_W(0 ~mg ~B)}\approx100$. The growth rate of in-situ boron coatings on PFCs was determined to be $\approx$1 nm/s per 10 mg/s during recent DIII-D experiments \cite{bortolon_observations_2020}. Wall conditioning matures following shots with a total of 706 mg BN powder injection and does not improve further following 68.5 mg Li powder injection. 
\section{Discussion} \label{sec:discussion}
Low-Z powder injection has been very suitable for reducing heat fluxes, increasing neutral build-up in the divertor, and improving wall conditions by forming low recycling coatings on the main plasma-facing components. Boron was suitable for cooling divertor heat fluxes and temperatures and showed the most significant improvement in wall conditions. However, it showed less effectiveness regarding divertor neutral compression. The neutral compression is higher and increases further with lithium and boron nitride injection. Moreover, the latter affect $Z_{eff}$ much less than boron. 54 mg/s BN reduced divertor electron temperature below to $\leq5$ eV. Very high rates of BN (204 mg/s) also substantially reduce ELM fluxes but cause losses of 24\% to $W_{mhd}$. 

The scenarios with the ion $B \times \nabla B$ drift pointing away from the SAS require lower powder injection rates to achieve substantial cooling effects and detachment. This is plausible since the onset of detachment also occurs at lower densities in this configuration \cite{guo_first_2019}.

Differences in divertor dissipation and core plasma dilution, particularly between boron and boron nitride, may be due to the different particle sizes and particle ablation. The B particles have a size up to 150 $\mu$m and can penetrate deeper into the plasma than the 65 $\mu$m BN particles. Experiments are planned to investigate the effect of particle size in comparison to modeling.

The line emission distributions obtained with tangential cameras are consistent with the temperature dependencies of the cooling potentials shown in figure \ref{fig:Figure1}(c) for the respective impurities. The data suggest that boron radiates at lower scrape-off layer temperatures than lithium, which may help optimize the SOL-to-core fraction in radiative losses.  

Future modeling will be necessary to disentangle the effects of particle size, injection rates, radiation behavior, and potentially drift effects on the divertor plasma distribution.

Powders injected at higher rates, i.e., several tens in case of large-grained boron and several hundreds of mg/s in case of fine-grained BN, increase the risk of the onset of tearing modes. These tearing modes are detrimental to confinement and occur during radiative mantle, and radiative divertor experiments with impurity gas seeding \cite{petrie_2019}. 

Wall conditions improved significantly after injecting powders into the closed divertor. The substantial reduction in recycling, carbon, and oxygen levels is consistent with results reported earlier for dedicated real-time wall conditioning experiments in lower single null configurations. Experiments with low recycling materials at ASDEX-U and DIII-D have shown that conditioning of material surfaces in the private flux region, the far SOL and components such as tungsten ion-cyclotron antennas is possible \cite{bortolon_real-time_2019, lunsford_2019, bortolon_observations_2020, krieger_psi_2022, effenberg_psi_2022}. ELMs and ballistic transport of neutral impurities may contribute to deposition on the main chamber wall.
An essential difference in the present work is that wall conditions improve by injecting boron powder into the divertor (compared to an upstream injection location in the plasma crown). This suggests that divertor powder injection may also be beneficial for reducing high-Z impurity sources in a new tungsten divertor \cite{abrams_2021}. 

A general advantage of powder injection is that certain impurities can be delivered in pure form (no dilution through additional hydrogen bonding, like methane or diborane) and are not toxic or explosive. Recent studies at EAST and KSTAR show promising results regarding real-time wall conditioning in long pulse scenarios \cite{xu_2020, gilson_wall_2021}. 
Results from ASDEX-U have also shown that boron nitride powder injection reduces the ammonia content by 90\% compared to nitrogen gas seeding \cite{lunsford_2019}. However, fuel (tritium) retention in low recycling coatings remains a concern, and future studies will be dedicated to this topic.

In most current experiments, the delivery of solid material used a dropping technique relying on gravitational acceleration. Top powder injection facilitates dissipation in upper divertor configuration but might be less effective in lower divertor configurations. Additional devices have been developed and used for horizontal material, and powder injection, also suitable for lower divertor applications \cite{vorenkamp_2017, nagy_2019, bortolon_2016, effenberg_psi_2022}. Bottom-up injection of materials into the lower divertor has recently been demonstrated at DIII-D and will be further developed. \cite{orlov_2021}.

\section{Summary} \label{sec:summary}

Injection of boron, boron nitride, and lithium powders into the closed small angle slot (SAS) divertor during high confinement plasma operation has been found to reduce the divertor heat fluxes and improve wall conditions at the DIII-D tokamak.

Boron radiates deeper in the SOL at lower temperatures (1-10 eV) while lithium radiates at higher electron temperatures near the separatrix (10-100 eV). Boron nitride combines the dissipative properties of both boron and nitrogen and appears to be most effective for divertor power dissipation. Li and BN powders have increased neutral compression and reduced heat fluxes more effectively than B powder. Divertor detachment was achieved at a BN rate of 54 mg/s (the minimum BN flow rate for detachment might be lower). BN at rates up to 204 mg/s also suppressed ELM-fluxes to the divertor but triggered $n=2$ tearing modes and reduced neutron rates and core energy confinement by 24\%. 

A total of 166 mg boron powder deposited throughout four discharges (corresponding to average rates on the order of $\approx$10 mg/s) improved the wall conditioning ratio of gas-input to electron density by a factor of $\approx$100. Low Z coatings growing on the PFCs at $\approx$1 nm/s per 10 mg/s of powders injected reduced wall recycling, carbon, and oxygen influxes.

Additional studies will be necessary to address the issue of fuel retention (tritium) and better understand the effects of particle size (44-150 $\mu$m), impurity species, and their impact on the plasma-surface interactions.

Real-time powder injection provides access to a more extensive range of impurity species for dissipative power exhaust. Optimization of the flow rates may achieve sufficient divertor dissipation while maintaining high core energy confinement. Low-Z powder injection into the divertor is a promising new approach to suppress high heat fluxes to the divertor targets and improve core-edge compatibility.
\section*{Acknowledgements}
This material is based upon work supported by the U.S. Department of Energy, Office of Science, Office of Fusion Energy Sciences, using the DIII-D National Fusion Facility, a DOE Office of Science user facility, under Awards DE-AC02-09CH11466 (PPPL), DE-FC02-04ER54698 (DIII-D), and DE-AC52-07NA27344 (LLNL).  The author Florian Effenberg gratefully acknowledges discussions with Galen Burke, Houyang Y. Guo, Adam McLean, Charles J. Lasnier, Jun Ren, Morgan W. Shafer, and Dinh Truong. The
DIII-D data shown in this paper can be obtained in digital format by following the links at \nolinkurl{https://fusion.gat.com/global/D3D_DMP}. The United States Government retains a non-exclusive, paid-up, irrevocable, world-wide license to publish or reproduce the published form of this manuscript, or allow others to do so, for United States Government purposes.
\section*{Disclaimer}
This report was prepared as an account of work sponsored by an agency of the United States Government. Neither the United States Government nor any agency thereof, nor any of their employees, makes any warranty, express or implied, or assumes any legal liability or responsibility for the accuracy, completeness, or usefulness of any information, apparatus, product, or process disclosed, or represents that its use would not infringe privately owned rights. Reference herein to any specific commercial product, process, or service by trade name, trademark, manufacturer, or otherwise does not necessarily constitute or imply its endorsement, recommendation, or favoring by the United States Government or any agency thereof. The views and opinions of authors expressed herein do not necessarily state or reflect those of the United States Government or any agency thereof.
\section*{References}

\bibliography{main}

\begin{thebibliography}{10}
\expandafter\ifx\csname url\endcsname\relax
  \def\url#1{\texttt{#1}}\fi
\expandafter\ifx\csname urlprefix\endcsname\relax\def\urlprefix{URL }\fi
\expandafter\ifx\csname href\endcsname\relax
  \def\href#1#2{#2} \def\path#1{#1}\fi

\bibitem{pitts_full_2013}
R.~A. Pitts, S.~Carpentier, F.~Escourbiac, T.~Hirai, V.~Komarov, S.~Lisgo,
  A.~S. Kukushkin, A.~Loarte, M.~Merola, A.~S. Naik, R.~Mitteau, M.~Sugihara,
  B.~Bazylev, P.~C. Stangeby, A full tungsten divertor for {ITER}: {Physics}
  issues and design status, Journal of Nuclear Materials 438 (2013) S48 -- S56.
\newblock \href {https://doi.org/10.1016/j.jnucmat.2013.01.008}
  {\path{doi:10.1016/j.jnucmat.2013.01.008}}.

\bibitem{national_academy_of_engineering_bringing_2021}
{National Academy of Engineering}, E.~National Academies~of Sciences,
  {and}~Medicine, Bringing {Fusion} to the {U}.{S}. {Grid}, The National
  Academies Press, Washington, DC, 2021.
\newblock \href {https://doi.org/10.17226/25991} {\path{doi:10.17226/25991}}.

\bibitem{maddison_moderation_2011}
G.~Maddison, C.~Giroud, G.~McCormick, J.~Alonso, B.~Alper, G.~Arnoux,
  P.~da~Silva Aresta~Belo, M.~Beurskens, A.~Boboc, S.~Brezinsek, I.~Coffey,
  S.~Devaux, T.~Eich, W.~Fundamenski, D.~Harting, A.~Huber, S.~Jachmich,
  I.~Jenkins, E.~Joffrin, M.~Kempenaars, M.~Lehnen, T.~Loarer, P.~Lomas,
  A.~Meigs, P.~Morgan, V.~Riccardo, F.~Rimini, M.~Stamp, G.~Telesca,
  H.~Thomsen, {JET EFDA contributors}, Moderation of divertor heat loads by
  fuelling and impurity seeding in well-confined {ELMy} {H}-mode plasmas on
  {JET}, Nuclear Fusion 51~(4) (2011) 042001.
\newblock \href {https://doi.org/10.1088/0029-5515/51/4/042001}
  {\path{doi:10.1088/0029-5515/51/4/042001}}.

\bibitem{samm_radiative_1993}
U.~Samm, G.~Bertschinger, P.~Bogen, J.~D. Hey, E.~Hintz, L.~Konen, Y.~T. Lie,
  A.~Pospieszczyk, D.~Rusbuldt, R.~P. Schorn, B.~Schweer, M.~Tokar,
  B.~Unterberg, Radiative edges under control by impurity fluxes, Plasma
  Physics and Controlled Fusion 35~(SB) (1993) B167--B175, publisher: IOP
  Publishing.
\newblock \href {https://doi.org/10.1088/0741-3335/35/sb/013}
  {\path{doi:10.1088/0741-3335/35/sb/013}}.

\bibitem{reinke_2011}
M.~Reinke, J.~Hughes, A.~Loarte, D.~Brunner, I.~Hutchinson, B.~LaBombard,
  J.~Payne, J.~Terry, Effect of {N2}, {Ne} and {Ar} seeding on {Alcator}
  {C-Mod} {H}-mode confinement, Journal of Nuclear Materials 415~(1,
  Supplement) (2011) S340--S344, {Proceedings} of the 19th International
  Conference on Plasma-Surface Interactions in Controlled Fusion.
\newblock \href {https://doi.org/https://doi.org/10.1016/j.jnucmat.2010.10.055}
  {\path{doi:https://doi.org/10.1016/j.jnucmat.2010.10.055}}.

\bibitem{kallenbach_partial_2015}
A.~Kallenbach, M.~Bernert, M.~Beurskens, L.~Casali, M.~Dunne, T.~Eich,
  L.~Giannone, A.~Herrmann, M.~Maraschek, S.~Potzel, F.~Reimold, V.~Rohde,
  J.~Schweinzer, E.~Viezzer, M.~Wischmeier, {the ASDEX Upgrade Team}, Partial
  detachment of high power discharges in {ASDEX} {Upgrade}, Nucl. Fusion 55~(5)
  (2015) 053026.
\newblock \href {https://doi.org/10.1088/0029-5515/55/5/053026}
  {\path{doi:10.1088/0029-5515/55/5/053026}}.

\bibitem{casali_improved_2020}
L.~Casali, T.~H. Osborne, B.~A. Grierson, A.~G. McLean, E.~T. Meier, J.~Ren,
  M.~W. Shafer, H.~Wang, J.~G. Watkins, Improved core-edge compatibility using
  impurity seeding in the small angle slot ({SAS}) divertor at {DIII}-{D},
  Physics of Plasmas 27~(6) (2020) 062506.
\newblock \href {https://doi.org/10.1063/1.5144693}
  {\path{doi:10.1063/1.5144693}}.

\bibitem{chen_2017}
J.~Chen, Y.~Duan, Z.~Yang, L.~Wang, K.~Wu, K.~Li, F.~Ding, H.~Mao, J.~Xu,
  W.~Gao, L.~Zhang, J.~Wu, G.-N. {Luo}, {EAST Team}, Radiative divertor
  behavior and physics in {Ar} seeded plasma on {EAST}, Chinese Physics B
  26~(9) (2017) 095205.
\newblock \href {https://doi.org/10.1088/1674-1056/26/9/095205}
  {\path{doi:10.1088/1674-1056/26/9/095205}}.

\bibitem{morisaki_radiated_2015}
T.~Morisaki, K.~Oyama, N.~Tamura, S.~Masuzaki, T.~Akiyama, G.~Motojima,
  J.~Miyazawa, B.~J. Peterson, N.~Ohno, H.~Yamada, {LHD Experiment Group},
  Radiated power distributions in impurity-seeded plasmas in {LHD}, Journal of
  Nuclear Materials 463 (2015) 640 -- 643.
\newblock \href {https://doi.org/10.1016/j.jnucmat.2015.01.016}
  {\path{doi:10.1016/j.jnucmat.2015.01.016}}.

\bibitem{effenberg_first_2019}
F.~Effenberg, S.~Brezinsek, Y.~Feng, R.~König, M.~Krychowiak, M.~Jakubowski,
  H.~Niemann, V.~Perseo, O.~Schmitz, D.~Zhang, T.~Barbui, C.~Biedermann,
  R.~Burhenn, B.~Buttenschön, G.~Kocsis, A.~Pavone, F.~Reimold, T.~Szepesi,
  H.~Frerichs, Y.~Gao, U.~Hergenhahn, S.~Kwak, M.~Otte, T.~{Sunn Pedersen},
  {W7-X Team}, First demonstration of radiative power exhaust with impurity
  seeding in the island divertor at {Wendelstein} 7-{X}, Nuclear Fusion 59~(10)
  (2019) 106020.
\newblock \href {https://doi.org/10.1088/1741-4326/ab32c4}
  {\path{doi:10.1088/1741-4326/ab32c4}}.

\bibitem{matthews_plasma_1995}
G.~F. Matthews, Plasma detachment from divertor targets and limiters, Journal
  of Nuclear Materials 220-222 (1995) 104 -- 116.
\newblock \href {https://doi.org/10.1016/0022-3115(94)00450-1}
  {\path{doi:10.1016/0022-3115(94)00450-1}}.

\bibitem{krasheninnikov_2016}
S.~I. Krasheninnikov, A.~S. Kukushkin, A.~A. Pshenov, Divertor plasma
  detachment, Physics of Plasmas 23~(5) (2016) 055602.
\newblock \href {https://doi.org/10.1063/1.4948273}
  {\path{doi:10.1063/1.4948273}}.

\bibitem{neuwirth_2012}
D.~Neuwirth, V.~Rohde, T.~{Schwarz-Selinger}, {ASDEX Upgrade Team}, Formation
  of ammonia during nitrogen-seeded discharges at {ASDEX} {Upgrade}, Plasma
  Physics and Controlled Fusion 54~(8) (2012) 085008.
\newblock \href {https://doi.org/10.1088/0741-3335/54/8/085008}
  {\path{doi:10.1088/0741-3335/54/8/085008}}.

\bibitem{laguardia_2017}
L.~Laguardia, R.~Caniello, A.~Cremona, G.~Gatto, G.~Gervasini, F.~Ghezzi,
  G.~Granucci, V.~Mellera, D.~Minelli, R.~Negrotti, M.~Pedroni, M.~Realini,
  D.~Ricci, N.~Rispoli, A.~Uccello, E.~Vassallo, Influence of {He} and {Ar}
  injection on ammonia production in {N2/D2} plasma in the medium flux {GyM}
  device, Nuclear Materials and Energy 12 (2017) 261--266, proceedings of the
  22nd International Conference on Plasma Surface Interactions 2016, 22nd PSI.
\newblock \href {https://doi.org/10.1016/j.nme.2017.05.009}
  {\path{doi:10.1016/j.nme.2017.05.009}}.

\bibitem{kallenbach_impurity_2013}
A.~Kallenbach, M.~Bernert, R.~Dux, L.~Casali, T.~Eich, L.~Giannone,
  A.~Herrmann, R.~McDermott, A.~Mlynek, H.~W. Müller, F.~Reimold,
  J.~Schweinzer, M.~Sertoli, G.~Tardini, W.~Treutterer, E.~Viezzer,
  R.~Wenninger, M.~Wischmeier, {ASDEX Upgrade Team}, Impurity seeding for
  tokamak power exhaust: from present devices via {ITER} to {DEMO}, Plasma
  Physics and Controlled Fusion 55~(12) (2013) 124041, publisher: IOP
  Publishing.
\newblock \href {https://doi.org/10.1088/0741-3335/55/12/124041}
  {\path{doi:10.1088/0741-3335/55/12/124041}}.

\bibitem{walker_neutron_2017}
R.~J. Walker, M.~R. Gilbert, Neutron activation of impurity seeding gases
  within a {DEMO} environment, Fusion Engineering and Design 124 (2017) 892 --
  895.
\newblock \href {https://doi.org/10.1016/j.fusengdes.2017.01.057}
  {\path{doi:10.1016/j.fusengdes.2017.01.057}}.

\bibitem{pigarov_radiative_2017}
A.~Y. Pigarov, Radiative detached divertor with acceptable separatrix {Zeff},
  Physics of Plasmas 24~(10) (2017) 102521.
\newblock \href {https://doi.org/10.1063/1.4986516}
  {\path{doi:10.1063/1.4986516}}.

\bibitem{effenberg_2020}
F.~Effenberg, A.~Bortolon, H.~Frerichs, B.~Grierson, J.~Lore, T.~Abrams,
  T.~Evans, Y.~Feng, R.~Lunsford, R.~Maingi, A.~Nagy, R.~Nazikian, D.~Orlov,
  J.~Ren, D.~Rudakov, W.~Wampler, H.~Wang, {3D} modeling of boron transport in
  {DIII-D} {L-mode} wall conditioning experiments, Nuclear Materials and Energy
  26 (2021) 100900.
\newblock \href {https://doi.org/10.1016/j.nme.2021.100900}
  {\path{doi:10.1016/j.nme.2021.100900}}.

\bibitem{poradzinski_2019}
M.~Poradziński, I.~Ivanova-Stanik, G.~Pełka, V.~P. Ridolfini, R.~Zagórski,
  Integrated power exhaust modelling for {DEMO} with lithium divertor, Fusion
  Engineering and Design 146 (2019) 1500--1504, sI:SOFT-30.
\newblock \href {https://doi.org/10.1016/j.fusengdes.2019.02.115}
  {\path{doi:10.1016/j.fusengdes.2019.02.115}}.

\bibitem{andruczyk_2020}
D.~Andruczyk, R.~Maingi, J.~S. Hu, G.~Z. Zuo, R.~Rizkallah, M.~Parsons,
  A.~Shone, D.~O'Dea, A.~Kapat, M.~Szott, S.~Stemmley, Z.~Sun, W.~Xu, X.~C.
  Meng, R.~Lunsford, E.~P. Gilson, A.~Diallo, K.~Tritz, {EAST Team}, Overview
  of lithium injection and flowing liquid lithium results from the {US}-{China}
  collaboration on {EAST}, Physica Scripta T171 (2020) 014067.
\newblock \href {https://doi.org/10.1088/1402-4896/ab6ce1}
  {\path{doi:10.1088/1402-4896/ab6ce1}}.

\bibitem{marenkov_2021}
E.~Marenkov, A.~Kukushkin, A.~Pshenov, Modeling the vapor shielding of a liquid
  lithium divertor target using {SOLPS} 4.3 code, Nuclear Fusion 61~(3) (2021)
  034001.
\newblock \href {https://doi.org/10.1088/1741-4326/abd642}
  {\path{doi:10.1088/1741-4326/abd642}}.

\bibitem{goldston_lithium_2016}
R.~J. Goldston, R.~Myers, J.~Schwartz, The lithium vapor box divertor, Phys.
  Scr. T167 (2016) 014017.
\newblock \href {https://doi.org/10.1088/0031-8949/T167/1/014017}
  {\path{doi:10.1088/0031-8949/T167/1/014017}}.

\bibitem{schwartz_2019}
J.~A. Schwartz, E.~D. Emdee, R.~Goldston, M.~Jaworski, Physics design for a
  lithium vapor box divertor experiment on {Magnum-PSI}, Nuclear Materials and
  Energy 18 (2019) 350--355.
\newblock \href {https://doi.org/10.1016/j.nme.2019.01.024}
  {\path{doi:10.1016/j.nme.2019.01.024}}.

\bibitem{snipes_1992}
J.~Snipes, E.~Marmar, J.~Terry, M.~Bell, R.~Budny, K.~Hill, D.~Jassby,
  D.~Mansfield, D.~Meade, H.~Park, J.~Strachan, B.~Stratton, E.~Synakowski,
  G.~Taylor, D.~Ruzic, M.~Shaheen, {TFTR Group}, Wall conditioning with
  impurity pellet injection on {TFTR}, Journal of Nuclear Materials 196-198
  (1992) 686--691.
\newblock \href {https://doi.org/10.1016/S0022-3115(06)80123-6}
  {\path{doi:10.1016/S0022-3115(06)80123-6}}.

\bibitem{sergeev_2012}
V.~Y. Sergeev, B.~V. Kuteev, A.~S. Bykov, S.~V. Krylov, V.~G. Skokov, V.~M.
  Timokhin, Lithium technologies for edge plasma control, Fusion Engineering
  and Design 87~(10) (2012) 1765--1769.
\newblock \href {https://doi.org/10.1016/j.fusengdes.2011.10.008}
  {\path{doi:10.1016/j.fusengdes.2011.10.008}}.

\bibitem{bortolon_real-time_2019}
A.~Bortolon, V.~Rohde, R.~Maingi, E.~Wolfrum, R.~Dux, A.~Herrmann, R.~Lunsford,
  R.~M. McDermott, A.~Nagy, A.~Kallenbach, D.~K. Mansfield, R.~Nazikian,
  R.~Neu, {ASDEX Upgrade Team}, Real-time wall conditioning by controlled
  injection of boron and boron nitride powder in full tungsten wall {ASDEX}
  {Upgrade}, Nuclear Materials and Energy 19 (2019) 384 -- 389.
\newblock \href {https://doi.org/10.1016/j.nme.2019.03.022}
  {\path{doi:10.1016/j.nme.2019.03.022}}.

\bibitem{bortolon_observations_2020}
A.~Bortolon, R.~Maingi, A.~Nagy, J.~Ren, J.~Duran, A.~Maan, D.~Donovan,
  J.~Boedo, D.~Rudakov, A.~Hyatt, T.~Wilks, M.~Shafer, C.~Samuell,
  M.~Fenstermacher, E.~Gilson, R.~Lunsford, D.~Mansfield, T.~Abrams,
  R.~Nazikian, Observations of wall conditioning by means of boron powder
  injection in {DIII-D} {H-mode} plasmas, Nuclear Fusion 60~(12) (2020) 126010.
\newblock \href {https://doi.org/10.1088/1741-4326/abaf31}
  {\path{doi:10.1088/1741-4326/abaf31}}.

\bibitem{gilson_wall_2021}
E.~Gilson, H.~Lee, A.~Bortolon, W.~Choe, A.~Diallo, S.~Hong, H.~Lee, J.~Lee,
  R.~Maingi, D.~Mansfield, A.~Nagy, S.~Park, I.~Song, J.~Song, S.~Yun, S.~Yoon,
  R.~Nazikian, Wall conditioning and {ELM} mitigation with boron nitride powder
  injection in {KSTAR}, Nuclear Materials and Energy 28 (2021) 101043.
\newblock \href {https://doi.org/10.1016/j.nme.2021.101043}
  {\path{doi:10.1016/j.nme.2021.101043}}.

\bibitem{sun_real_2019}
Z.~Sun, R.~Maingi, J.~Hu, W.~Xu, G.~Zuo, Y.~Yu, C.~Wu, M.~Huang, X.~Meng,
  L.~Zhang, L.~Wang, S.~Mao, F.~Ding, D.~Mansfield, J.~Canik, R.~Lunsford,
  A.~Bortolon, X.~Gong, {EAST Team}, Real time wall conditioning with lithium
  powder injection in long pulse {H}-mode plasmas in {EAST} with tungsten
  divertor, Nuclear Materials and Energy 19 (2019) 124--130.
\newblock \href {https://doi.org/10.1016/j.nme.2019.02.029}
  {\path{doi:10.1016/j.nme.2019.02.029}}.

\bibitem{maingi_elm_2018}
R.~Maingi, J.~S. Hu, Z.~Sun, K.~Tritz, G.~Z. Zuo, W.~Xu, M.~Huang, X.~C. Meng,
  J.~M. Canik, A.~Diallo, R.~Lunsford, D.~K. Mansfield, T.~H. Osborne, X.~Z.
  Gong, Y.~F. Wang, Y.~Y. Li, {EAST Team}, {ELM} elimination with {Li} powder
  injection in {EAST} discharges using the tungsten upper divertor, Nuclear
  Fusion 58~(2) (2018) 024003, publisher: IOP Publishing.
\newblock \href {https://doi.org/10.1088/1741-4326/aa9e3f}
  {\path{doi:10.1088/1741-4326/aa9e3f}}.

\bibitem{sun_suppression_2021}
Z.~Sun, A.~Diallo, R.~Maingi, Y.~Qian, K.~Tritz, Y.~Wang, Y.~Wang, A.~Bortolon,
  A.~Nagy, L.~Zhang, Y.~Duan, Y.~Ye, H.~Zhao, H.~Wang, X.~Gu, G.~Zuo, W.~Xu,
  M.~Huang, C.~Li, X.~Meng, C.~Zhou, H.~Liu, Q.~Zang, L.~Wang, J.~Qian, G.~Xu,
  X.~Gong, J.~Hu, {EAST Team}, Suppression of edge localized modes with
  real-time boron injection using the tungsten divertor in {EAST}, Nucl. Fusion
  61~(1) (2021) 014002.
\newblock \href {https://doi.org/10.1088/1741-4326/abc763}
  {\path{doi:10.1088/1741-4326/abc763}}.

\bibitem{nespoli_2020}
F.~Nespoli, N.~Ashikawa, E.~Gilson, R.~Lunsford, S.~Masuzaki, M.~Shoji,
  T.~Oishi, C.~Suzuki, A.~Nagy, A.~Mollen, N.~Pablant, K.~Ida, M.~Yoshinuma,
  N.~Tamura, D.~Gates, T.~Morisaki, {LHD experiment group}, First impurity
  powder injection experiments in {LHD}, Nuclear Materials and Energy 25 (2020)
  100842.
\newblock \href {https://doi.org/10.1016/j.nme.2020.100842}
  {\path{doi:10.1016/j.nme.2020.100842}}.

\bibitem{lunsford_characterization_2021}
R.~Lunsford, C.~Killer, A.~Nagy, D.~A. Gates, T.~Klinger, A.~Dinklage,
  G.~Satheeswaran, G.~Kocsis, S.~A. Lazerson, F.~Nespoli, N.~A. Pablant, A.~von
  Stechow, A.~Alonso, T.~Andreeva, M.~Beurskens, C.~Biedermann, S.~Brezinsek,
  K.~J. Brunner, B.~Buttenschön, D.~Carralero, G.~Cseh, P.~Drewelow,
  F.~Effenberg, T.~Estrada, O.~P. Ford, O.~Grulke, U.~Hergenhahn, U.~Höfel,
  J.~Knauer, M.~Krause, M.~Krychowiak, S.~Kwak, A.~Langenberg, U.~Neuner,
  D.~Nicolai, A.~Pavone, A.~Puig~Sitjes, K.~Rahbarnia, J.~Schilling,
  J.~Svensson, T.~Szepesi, H.~Thomsen, T.~Wauters, T.~Windisch, V.~Winters,
  D.~Zhang, L.~Zsuga, {W7-X Team}, Characterization of injection and
  confinement improvement through impurity induced profile modifications on the
  {Wendelstein 7-X} stellarator, Physics of Plasmas 28~(8) (2021) 082506.
\newblock \href {https://doi.org/10.1063/5.0047274}
  {\path{doi:10.1063/5.0047274}}.

\bibitem{stangeby_2022}
P.~C. Stangeby, E.~A. Unterberg, J.~W. Davis, T.~Abrams, A.~Bortolon, I.~Bykov,
  D.~Donovan, H.~Y. Guo, R.~Kolasinski, A.~W. Leonard, J.~H. Nichols, D.~L.
  Rudakov, G.~Sinclair, D.~M. Thomas, J.~G. Watkins, Developing solid-surface
  plasma facing components for pilot plants and reactors with replenishable
  wall claddings and continuous surface conditioning. part a: concepts and
  questions, Plasma Physics and Controlled Fusion 64~(5) (2022) 055018.
\newblock \href {https://doi.org/10.1088/1361-6587/ac5a7c}
  {\path{doi:10.1088/1361-6587/ac5a7c}}.

\bibitem{abrams_2017}
T.~Abrams, R.~Ding, H.~Guo, D.~Thomas, C.~Chrobak, D.~Rudakov, A.~McLean,
  E.~Unterberg, A.~Briesemeister, P.~Stangeby, J.~Elder, W.~Wampler,
  J.~Watkins, The inter-{ELM} tungsten erosion profile in {DIII-D} {H-mode}
  discharges and benchmarking with {ERO}$+${OEDGE} modeling, Nuclear Fusion
  57~(5) (2017) 056034.
\newblock \href {https://doi.org/10.1088/1741-4326/aa66b2}
  {\path{doi:10.1088/1741-4326/aa66b2}}.

\bibitem{fenstermacher_tangentially_1997}
M.~E. Fenstermacher, W.~H. Meyer, R.~D. Wood, D.~G. Nilson, R.~Ellis, N.~H.
  Brooks, A tangentially viewing visible {TV} system for the {DIII}-{D}
  divertor, Review of Scientific Instruments 68~(1) (1997) 974--977.
\newblock \href {https://doi.org/10.1063/1.1147729}
  {\path{doi:10.1063/1.1147729}}.

\bibitem{brooks_1992}
N.~H. Brooks, A.~Howald, K.~Klepper, P.~West, Multichord spectroscopy of the
  {DIII‐D} divertor region, Review of Scientific Instruments 63~(10) (1992)
  5167--5169.
\newblock \href {https://doi.org/10.1063/1.1143469}
  {\path{doi:10.1063/1.1143469}}.

\bibitem{morisaki_2013}
T.~Morisaki, S.~Masuzaki, M.~Kobayashi, M.~Shoji, J.~Miyazawa, R.~Sakamoto,
  G.~Motojima, M.~Goto, H.~Funaba, H.~Tanaka, K.~Tanaka, I.~Yamada, S.~Ohdachi,
  H.~Yamada, A.~Komori, {LHD Experiment Group}, Initial experiments towards
  edge plasma control with a closed helical divertor in {LHD}, Nuclear Fusion
  53~(6) (2013) 063014.
\newblock \href {https://doi.org/10.1088/0029-5515/53/6/063014}
  {\path{doi:10.1088/0029-5515/53/6/063014}}.

\bibitem{guo_first_2019}
H.~Guo, H.~Wang, J.~Watkins, L.~Casali, B.~Covele, A.~Moser, T.~Osborne,
  C.~Samuell, M.~Shafer, P.~Stangeby, D.~Thomas, J.~Boedo, R.~Buttery,
  R.~Groebner, D.~Hill, L.~Holland, A.~Hyatt, A.~Jaervinen, A.~Kellman, L.~Lao,
  C.~Lasnier, A.~Leonard, C.~Murphy, J.~Ren, C.~Sang, A.~Sontag, T.~Taylor,
  {the DIII-D Team}, First experimental tests of a new small angle slot
  divertor on {DIII}-{D}, Nucl. Fusion 59~(8) (2019) 086054.
\newblock \href {https://doi.org/10.1088/1741-4326/ab26ee}
  {\path{doi:10.1088/1741-4326/ab26ee}}.

\bibitem{shafer_dependence_2019}
M.~Shafer, B.~Covele, J.~Canik, L.~Casali, H.~Guo, A.~Leonard, J.~Lore,
  A.~McLean, A.~Moser, P.~Stangeby, D.~Taussig, H.~Wang, J.~Watkins, Dependence
  of neutral pressure on detachment in the small angle slot divertor at
  {DIII}-{D}, Nuclear Materials and Energy 19 (2019) 487--492.
\newblock \href {https://doi.org/10.1016/j.nme.2019.04.003}
  {\path{doi:10.1016/j.nme.2019.04.003}}.

\bibitem{casali_2019}
L.~Casali, B.~Covele, H.~Guo, The effect of neutrals in the new {SAS} divertor
  at {DIII-D} as modelled by {SOLPS}, Nuclear Materials and Energy 19 (2019)
  537--543.
\newblock \href {https://doi.org/10.1016/j.nme.2019.03.021}
  {\path{doi:10.1016/j.nme.2019.03.021}}.

\bibitem{casali_2021}
L.~Casali, D.~Eldon, A.~McLean, T.~Osborne, A.~Leonard, B.~Grierson, J.~Ren,
  Impurity leakage and radiative cooling in the first nitrogen and neon seeding
  study in the closed {DIII-D} {SAS} configuration, Nuclear Fusion 62~(2)
  (2022) 026021.
\newblock \href {https://doi.org/10.1088/1741-4326/ac3e84}
  {\path{doi:10.1088/1741-4326/ac3e84}}.

\bibitem{fevrier_2021}
O.~Février, H.~Reimerdes, C.~Theiler, D.~Brida, C.~Colandrea, H.~{De
  Oliveira}, B.~Duval, D.~Galassi, S.~Gorno, S.~Henderson, M.~Komm, B.~Labit,
  B.~Linehan, L.~Martinelli, A.~Perek, H.~Raj, U.~Sheikh, C.~Tsui, M.~Wensing,
  Divertor closure effects on the {TCV} boundary plasma, Nuclear Materials and
  Energy 27 (2021) 100977.
\newblock \href {https://doi.org/10.1016/j.nme.2021.100977}
  {\path{doi:10.1016/j.nme.2021.100977}}.

\bibitem{kallenbach_2019}
A.~Kallenbach, M.~Balden, R.~Dux, T.~Eich, C.~Giroud, A.~Huber, G.~Maddison,
  M.~Mayer, K.~McCormick, R.~Neu, T.~Petrie, T.~Pütterich, J.~Rapp, M.~Reinke,
  K.~Schmid, J.~Schweinzer, S.~Wolfe, {ASDEX Upgrade Team}, {DIII-D Team},
  {Alcator Team}, {JET-EFDA Contributors}, Plasma surface interactions in
  impurity seeded plasmas, Journal of Nuclear Materials 415~(1, Supplement)
  (2011) S19--S26, proceedings of the 19th International Conference on
  Plasma-Surface Interactions in Controlled Fusion.
\newblock \href {https://doi.org/10.1016/j.jnucmat.2010.11.105}
  {\path{doi:10.1016/j.jnucmat.2010.11.105}}.

\bibitem{carolan_1983}
P.~G. Carolan, V.~A. Piotrowicz, The behaviour of impurities out of coronal
  equilibrium, Plasma Physics 25~(10) (1983) 1065--1086.
\newblock \href {https://doi.org/10.1088/0032-1028/25/10/001}
  {\path{doi:10.1088/0032-1028/25/10/001}}.

\bibitem{summers_atomic_2002}
H.~P. Summers, N.~R. Badnell, M.~G. O{\textquotesingle}Mullane, A.~D.
  Whiteford, R.~Bingham, B.~J. Kellett, J.~Lang, K.~H. Behringer, U.~Fantz,
  K.-D. Zastrow, S.~D. Loch, M.~S. Pindzola, D.~C. Griffin, C.~P. Ballance,
  Atomic data for modelling fusion and astrophysical plasmas, Plasma Physics
  and Controlled Fusion 44~(12B) (2002) B323--B338.
\newblock \href {https://doi.org/10.1088/0741-3335/44/12b/323}
  {\path{doi:10.1088/0741-3335/44/12b/323}}.

\bibitem{nagy_multi-species_2018}
A.~Nagy, A.~Bortolon, D.~M. Mauzey, E.~Wolfe, E.~P. Gilson, R.~Lunsford,
  R.~Maingi, D.~K. Mansfield, R.~Nazikian, A.~L. Roquemore, A multi-species
  powder dropper for magnetic fusion applications, Review of Scientific
  Instruments 89~(10) (2018) 10K121.
\newblock \href {https://doi.org/10.1063/1.5039345}
  {\path{doi:10.1063/1.5039345}}.

\bibitem{heidbrink_1997}
W.~W. Heidbrink, P.~L. Taylor, J.~A. Phillips, Measurements of the neutron
  source strength at {DIII-D}, Review of Scientific Instruments 68~(1) (1997)
  536--539.
\newblock \href {https://doi.org/10.1063/1.1147646}
  {\path{doi:10.1063/1.1147646}}.

\bibitem{petrie_2019}
T.~Petrie, B.~Grierson, T.~Osborne, C.~Petty, F.~Turco, S.~Allen,
  M.~Fenstermacher, J.~Ferron, H.~Guo, E.~Hinson, R.~L. Haye, C.~Lasnier,
  A.~Leonard, A.~McLean, B.~Victor, H.~Wang, J.~Watkins, High performance
  double-null plasmas under radiating divertor and mantle scenarios on
  {DIII-D}, Nuclear Fusion 59~(8) (2019) 086053.
\newblock \href {https://doi.org/10.1088/1741-4326/ab2936}
  {\path{doi:10.1088/1741-4326/ab2936}}.

\bibitem{lunsford_2019}
R.~Lunsford, V.~Rohde, A.~Bortolon, R.~Dux, A.~Herrmann, A.~Kallenbach,
  R.~McDermott, P.~David, A.~Drenik, F.~Laggner, R.~Maingi, D.~K. Mansfield,
  A.~Nagy, R.~Neu, E.~Wolfrum, {ASDEX Upgrade Team}, Active conditioning of
  {ASDEX} upgrade tungsten plasma-facing components and discharge enhancement
  through boron and boron nitride particulate injection, Nuclear Fusion 59~(12)
  (2019) 126034.
\newblock \href {https://doi.org/10.1088/1741-4326/ab4095}
  {\path{doi:10.1088/1741-4326/ab4095}}.

\bibitem{krieger_psi_2022}
K.~Krieger, M.~Balden, A.~Bortolon, F.~Laggner, V.~Rohde, W.~Wampler, {ASDEX
  Upgrade Team}, Boron redistribution after boron powder injection in {ASDEX}
  {Upgrade}, 2022, 25th International Conference on Plasma Surface Interactions
  2022, 25th PSI.

\bibitem{effenberg_psi_2022}
F.~Effenberg, S.~Abe, F.~Laggner, D.~Rudakov, G.~Sinclair, W.~W.R., T.~Abrams,
  D.~Mauzey, A.~Nagy, F.~Scotti, R.~Nazikian, {DIII-D Team}, In-situ growth of
  silicon-rich layers with {Si} granule injection in {DIII-D} {H-mode} plasmas,
  2022, 25th International Conference on Plasma Surface Interactions 2022, 25th
  PSI.

\bibitem{abrams_2021}
T.~Abrams, G.~Sinclair, J.~H. Nichols, E.~A. Unterberg, D.~C. Donovan,
  J.~Duran, J.~D. Elder, F.~Glass, B.~A. Grierson, H.~Y. Guo, T.~Hall, X.~Ma,
  R.~Maurizio, A.~G. McLean, C.~Murphy, R.~Nguyen, D.~L. Rudakov, P.~C.
  Stangeby, D.~M. Thomas, S.~A. Zamperini, Design and physics basis for the
  upcoming {DIII-D} {SAS}-{VW} campaign to quantify tungsten leakage and
  transport in a new slot divertor geometry, Physica Scripta 96~(12) (2021)
  124073.
\newblock \href {https://doi.org/10.1088/1402-4896/ac3c5f}
  {\path{doi:10.1088/1402-4896/ac3c5f}}.

\bibitem{xu_2020}
W.~Xu, J.~S. Hu, Z.~Sun, R.~Maingi, L.~Zhang, Y.~W. Yu, C.~L. Li, G.~Z. Zuo,
  Y.~Z. Qian, M.~Huang, X.~C. Meng, W.~Gao, Y.~M. Duan, Y.~J. Chen, K.~Wang,
  X.~D. Lin, X.~Gao, Effect of lithium coating on long pulse high performance
  plasma discharges in {EAST}, Plasma Physics and Controlled Fusion 62~(8)
  (2020) 085012.
\newblock \href {https://doi.org/10.1088/1361-6587/ab9b3a}
  {\path{doi:10.1088/1361-6587/ab9b3a}}.

\bibitem{vorenkamp_2017}
M.~S. Vorenkamp, A.~Nagy, A.~Bortolon, R.~Lunsford, R.~Maingi, D.~K. Mansfield,
  A.~L. Roquemore, Recent upgrades of the {DIII-D} impurity granule injector,
  Fusion Science and Technology 72~(3) (2017) 488--495.
\newblock \href {https://doi.org/10.1080/15361055.2017.1335144}
  {\path{doi:10.1080/15361055.2017.1335144}}.

\bibitem{nagy_2019}
A.~Nagy, A.~Bortolon, D.~Gates, E.~Gilson, C.~Killer, T.~Klinger, R.~Lunsford,
  R.~Maingi, D.~Mansfield, D.~Mauzey, R.~Nazikian, L.~Roquemore, E.~Wolfe, A
  horizontal powder injector for {W7-X}, Fusion Engineering and Design 146
  (2019) 1403--1407.
\newblock \href {https://doi.org/10.1016/j.fusengdes.2018.12.099}
  {\path{doi:10.1016/j.fusengdes.2018.12.099}}.

\bibitem{bortolon_2016}
A.~Bortolon, R.~Maingi, D.~Mansfield, A.~Nagy, A.~Roquemore, L.~Baylor,
  N.~Commaux, G.~Jackson, E.~Gilson, R.~Lunsford, P.~Parks, C.~Chrystal,
  B.~Grierson, R.~Groebner, S.~Haskey, M.~Makowski, C.~Lasnier, R.~Nazikian,
  T.~Osborne, D.~Shiraki, M.~V. Zeeland, High frequency pacing of edge
  localized modes by injection of lithium granules in {DIII-D} {H-mode}
  discharges, Nuclear Fusion 56~(5) (2016) 056008.
\newblock \href {https://doi.org/10.1088/0029-5515/56/5/056008}
  {\path{doi:10.1088/0029-5515/56/5/056008}}.

\bibitem{orlov_2021}
D.~{Orlov}, {Hypervelocity impact in stellar media: Spacecraft Heat Shield
  study in DIII-D*}, in: APS Division of Plasma Physics Meeting Abstracts, Vol.
  2021 of APS Meeting Abstracts, 2021, p. WI02.002.

\end{thebibliography}

\end{document}